# Hybrid straintronics and voltage-controlled-magnetic-anisotropy: Precessional switching of a perpendicular anisotropy magneto-tunneling junction without a magnetic field

Justine L. Drobitch, Md Ahsanul Abeed and Supriyo Bandyopadhyay[1]

Department of Electrical and Computer Engineering

Virginia Commonwealth University, Richmond, VA 23284, USA

Abstract:

We propose an *all-electric* implementation of a precessionally switched perpendicular magnetic anisotropy magneto-tunneling-junction (p-MTJ) based toggle memory cell where data is written with voltage-controlled-magnetic-anisotropy (VCMA) without requiring an in-plane magnetic field. This is achieved by fashioning the soft layer of the MTJ out of a two-phase (magnetostrictive/piezoelectric) multiferroic which is electrically stressed to produce an *effective* in-plane magnetic field around which the magnetization precesses to complete a flip. The VCMA voltage pulse duration and the stress duration are adjusted to obtain a high switching probability. A two-terminal energy-efficient cell, that is compatible with crossbar architecture and high cell density, is designed.

[1]Corresponding author. Email: sbandy@vcu.edu

Voltage-switched perpendicular-magnetic-anisotropy magneto-tunneling-junctions (p-MTJ) are the preferred embodiments of energy-efficient, non-volatile, high-density magnetic random access memory (MRAM) cells. Their resistances can be switched with a voltage inducing either strain [1, 2] or magnetic anisotropy change (voltage controlled magnetic anisotropy: VCMA) [3-6] in the soft layer of the p-MTJ to write the binary bit 0 or 1. The use of voltage to switch, as opposed to spin current used in traditional spin-transfer torque and domain wall motion switching, reduces the write energy dissipation.

Fig. 1 shows a typical VCMA-switched p-MTJ where the soft and hard ferromagnetic layers have perpendicular magnetic anisotropy. A voltage $V_{VCMA}$ dropped across the spacer layer injects electrons into the insulating spacer layer. These charges accumulate at the interface between the spacer and the soft layer, modifying the band structure and the occupation of the bands in the latter. That changes the magnetic anisotropy of the soft layer from "perpendicular" to "in-plane" [7, 8] and can rotate the easy axis of magnetization from "out-of-plane" to "in-plane", resulting in $90^0$ rotation of the soft layer's magnetization vector as shown by the solid arrows in Fig. 1(a).

The $90^0$ rotation, however, does not result in complete magnetization reversal of the soft layer and produces low tunneling magnetoresistance ratio (TMR) of the p-MTJ, which inhibits unambiguous reading of the bits stored in the resistance state of the p-MTJ. To overcome this drawback, an in-plane magnetic field is applied as shown in Fig. 1(b) to bring about the desired $180^0$ switching [4, 9, 10]. Once the magnetization vector is dislodged from the stable out-of-plane direction by $V_{VCMA}$, it begins to precess about the magnetic field. By adjusting the $V_{VCMA}$ voltage pulse width to approximately one-half of the precessional period, the magnetization vector can be made to undergo ~$180^0$ rotation and thus flip [4]. The probability of flipping in this fashion depends on the in-plane magnetic field strength and the voltage pulse duration when room temperature thermal noise is present. For certain combinations of the above two parameters, the probability can exceed 90% [11, 12]. This probability may be adequate for memory applications where various on-chip error detection and correction schemes are available. There are numerous reports of VCMA switching [13-16] and recent progress has been reviewed in refs. [17, 18].

The major disadvantage of VCMA-switched p-MTJs is the need for the in-plane magnetic field which is an inconvenience. Any scheme that results in all-electric switching *without any magnetic field* will be preferred, and in this paper we present such a scheme. What inspires this scheme is the realization that there is no need to have an actual magnetic field to induce precessional motion of the magnetization vector. Any *effective* magnetic field will do [10]. Such an effective field can be generated in the soft layer by spin orbit interaction [19] or mechanical strain if the soft layer is magnetostrictive [20]. The strain can be generated electrically, resulting in elimination of the magnetic field, and an all-electric rendition.

Let us focus on the strain-based approach. The in-plane cross-section of a p-MTJ can be made elliptical or circular. Here, we will consider a slightly elliptical cross-section, but we have verified that the results are qualitatively unchanged if the cross-section is circular. Biaxial strain can be generated within the soft layer in the following way. The p-MTJ stack is fabricated on a poled piezoelectric thin film with the soft layer in contact with the film. This configuration is shown in Fig. 2. A voltage is applied over the piezoelectric film via electrodes on the surface. These electrodes are arranged in a suitable pattern to generate biaxial strain in the region of the piezoelectric underneath the soft layer (compressive strain along the major axis and tensile strain along the minor axis, or vice versa, depending on the polarity of the voltage) [21-23]. The strain is partially or fully transferred to the soft layer through any ultrathin adhesion layer as shown in Fig. 2. If the soft layer is magnetostrictive, then the strain acts as an effective in-plane magnetic field within the soft layer.

If the soft layer's magnetostriction coefficient is positive (e.g. in materials like Terfenol-D or Galfenol) then compressive stress along the major axis results in an effective magnetic field along the minor axis, while tensile strain results in an effective magnetic field along the major axis. Similarly, compressive strain along the minor axis will result in an effective magnetic field along the major axis and tensile strain will result in an effective magnetic field along the minor axis. The situation will be the exact opposite if the magnetostriction coefficient of the soft layer is negative (e.g. in Co, Ni). This is depicted in Fig. 3. Thus, the biaxial strain will always produce an effective magnetic field that is directed along either the major or the minor axis of the ellipse. This is the effective magnetic field around which the magnetization vector precesses when a VCMA voltage dislodges it from the out-of-plane orientation.

To model the precessional magneto-dynamics, we consider the potential energy of the elliptical soft layer which depends on the orientation of its magnetization vector. We assume that the $z$-axis is aligned along the major axis of the ellipse and that $\theta$ and $\phi$ are the polar and azimuthal angles of the magnetization vector (see Fig. 2). The potential energy is given by

$$U = \underbrace{\frac{\mu_0}{2} M_s^2 \Omega \left[ N_{d-zz} \cos^2\theta + N_{d-yy} \sin^2\theta \sin^2\phi + N_{d-xx} \sin^2\theta \cos^2\phi \right]}_{\text{shape anisotropy energy}} \\ \underbrace{- \left[ K_{s0} \frac{\Omega}{d} + C \frac{V_{VCMA}}{d \times t_b} \Omega \right] \sin^2\theta \cos^2\phi}_{\text{surface anisotropy energy}} + \underbrace{U_{stress}}_{\text{stress anisotropy energy}},$$ (1)

where $\mu_0$ is the permeability of free space, $M_s$ is the saturation magnetization of the soft layer, $\Omega$ is the soft layer's volume, and $U_{stress}$ is the stress anisotropy energy. The quantity $K_{s0}$ is the intrinsic surface anisotropy energy per unit lateral area, $C$ is the VCMA constant and $t_b$ is the spacer layer thickness. The quantities $N_{d-mm}$ are the demagnetization constants along the coordinate axes $\left(N_{d-xx} + N_{d-yy} + N_{d-zz} = 1\right)$. They depend on the dimensions of the major axis, the minor axis and the thickness ($d$) of the soft layer. When uniaxial stress is present along the major axis of the soft layer, $U_{stress} = -(3/2)\lambda\sigma\cos^2\theta$ and when it is present along the minor axis, $U_{stress} = -(3/2)\lambda\sigma\sin^2\theta\sin^2\phi$. Here, $\lambda$ is the magnetostriction coefficient and $\sigma$ is the stress (positive for tensile and negative for compressive).

Let us focus on the stress anisotropy term $U_{stress}$. If uniaxial stress has been applied along the major axis, then, as long as the $\lambda\sigma$ product is positive, $U_{stress}$ is minimized when $\theta = 0^0$ or $180^0$, which means that stress would prefer to align the magnetization vector (or equivalently the easy axis) along the major axis of the ellipse. On the other hand, if the $\lambda\sigma$ product is negative, then $U_{stress}$ is minimized when $\theta = \phi = 90^0$ and stress would prefer to align the easy axis along the minor axis of the ellipse. Similar situations will arise for uniaxial stress applied along the minor axis. Overall, we can view this as if stress results in an effective magnetic field in the direction of the preferred easy axis and that is why the magnetization vector tends to align along the easy axis [19]. This is the basis of Fig. 3. The magnitude of this effective magnetic field is $\left|3\lambda\sigma/(\mu_0 M_s)\right|$ [19]. For a highly magnetostrictive nanomagnet made of Terfenol-D $Tb_{0.27}Dy_{0.73}Fe_{1.95}$ ($\lambda = 600$ ppm, $M_s = 8\times10^5$ A/m) [24], 50 MPa uniaxial stress will generate an effective magnetic field of strength $H_{eff}$ = 90,000 A/m (1,125 Oe).

Soft layers of p-MTJs are designed with large $K_{s0}$ and are very thin. As a result, when $V_{VCMA} = 0$, the surface anisotropy term is dominant in Equation (1) and the magnetization will point out of plane because the total energy $U$ is minimized when $\theta = 90^0$ and $\phi = 0^0$ or $180^0$. We can progressively decrease the surface anisotropy by applying a negative $V_{VCMA}$ (for a positive $C$) and ultimately the magnetization vector will prefer to rotate to in-plane orientation because the easy axis will lie in the plane of the soft layer when the surface anisotropy term loses dominance. This corresponds to the $90^0$ switching situation shown in Fig. 1(a). However, if uniaxial stress is present along a principal axis of the elliptical soft layer, then there will be an effective magnetic field along the same or the other principal axis. The magnetization vector will precess about this effective magnetic field when it comes out of the out-of-plane orientation and that will take it past the in-plane orientation (rotation > $90^0$). By adjusting the $V_{VCMA}$ pulse width to approximately one-half of the precessional period, we can make the magnetization vector undergo a complete $180^0$ rotation, resulting in a flip, just as in refs. [4, 9, 10]. Note that the effective magnetic field in this case is generated via *electrically* produced stress and hence there is *no actual magnetic field*. Therefore, this is an all-electric implementation.

To verify that this switching paradigm can work, we have modeled the precessional dynamics in the presence of thermal noise. For this, we have solved the stochastic Landau-Lifshitz-Gilbert (s-LLG) equation within the macrospin approximation. The s-LLG equation can be reduced to two uncoupled equations for the polar and azimuthal angles of the magnetization vector [25]:

$$\left(1+\alpha^2\right)\frac{d\theta(t)}{dt} = \frac{\gamma}{M_V}\left[B_{0e}(t)\sin\theta(t) - 2\alpha B(t)\sin\theta(t)\cos\theta(t) + \alpha P_\theta(t) + P_\phi(t)\right]$$
$$\left(1+\alpha^2\right)\frac{d\phi(t)}{dt} = \frac{\gamma}{M_V}\left[\alpha B_{0e}(t) + 2B(t)\cos\theta(t) - \left(1/\sin\theta(t)\right)\left\{P_\theta(t) - \alpha P_\phi(t)\right\}\right]$$
(2)

Here,

$$B(t) = \frac{1}{2}\mu_0 M_s^2 \Omega\left[N_{d-xx}\cos^2\phi(t) + N_{d-yy}\sin^2\phi(t) - N_{d-zz}\right] - \left(K_{s0}\frac{\Omega}{l} + C\Omega\frac{V_{bias}}{t\times t_b}\right)\cos^2\phi(t)$$
$$\quad + \frac{3}{2}\lambda_s\sigma\Omega$$

$$B_{0e}(t) = \left\{\frac{1}{2}\mu_0 M_s^2 \Omega\left(N_{d-xx} - N_{d-yy}\right) - \left(K_{s0}\frac{\Omega}{l} + C\Omega\frac{V_{bias}}{t\times t_b}\right)\right\}\sin(2\phi(t))$$
(3)

$$M_V = \mu_0 M_s \Omega$$

$$P_\theta(t) = M_V\left[h_x(t)\cos\theta(t)\cos\phi(t) + h_y(t)\cos\theta(t)\sin\phi(t) - h_z(t)\sin\theta(t)\right]$$

$$P_\phi(t) = M_V\left[h_y(t)\cos\phi(t) - h_x(t)\sin\phi(t)\right]$$

The quantity $\gamma = \frac{2\mu_B \mu_0}{\hbar} = 2.21 \times 10^5 \, (\text{rad.m})/(\text{A.s})$ is a universal constant and $\alpha$ is the Gilbert damping constant. The quantities $h_x(t), h_y(t), h_z(t)$ are three independent (uncorrelated) random magnetic fields given by

$$h_i(t) = \sqrt{\frac{2\alpha kT}{\gamma(1+\alpha^2)M_V \Delta t}} G_{(0,1)}(t) \qquad (i = x, y, z) \tag{4}$$

where $\Delta t$ is the inverse of the attempt frequency (it is also the time step used in the simulation), $k$ is the Boltzmann constant, $T$ is the absolute temperature, and $G_{(0,1)}(t)$ is a Gaussian with zero mean and unit standard deviation. The simulation method is described in numerous publications [25-28] and hence not repeated here. The various material parameters and other constants used in the simulations are listed in Table I along with references from which these parameter values are adopted.

We assume that the soft layer is made of Terfenol-D (which has a positive magnetostriction coefficient). At time $t = 0$, the magnetization points up out of the soft layer's plane owing to perpendicular magnetic anisotropy (surface anisotropy term dominant), i.e. $m_x = +1$, $m_y = m_z = 0$, or $\theta = 90^0$, $\phi = 0^0$. Then, at $t = 0$, we turn on $V_{VCMA}$ (= -629.5 mV) along with a uniaxial compressive stress of 50 MPa along the major axis of the soft layer ($z$-axis). This stress generates an effective magnetic field of $\sim 9 \times 10^4$ A/m along the minor axis ($y$-axis) around which the magnetization of the soft layer will precess. We then follow the evolution of $\theta$, $\phi$ and $m_x$ with time. We generate 1,000 switching trajectories and plot $\theta(t)$, $\phi(t)$ and $m_x(t)$ (for each trajectory) in Figs. 4(a), (b) and (c). Note from Fig. 4(c) that $m_x$ executes damped oscillation between +1 and -1 and ultimately approaches 0, indicating that the magnetization vector will come to rest along the direction of the effective magnetic field (due to stress), which is along the $y$-axis ($m_y = +1$, $m_x = m_z = 0$). This is evident from the fact that $\theta$ and $\phi$ both approach $90^0$ with time. The magnetization vector executes damped precession around the effective magnetic field and finally comes to rest along the field because of Gilbert damping. Owing to thermal noise, there is a spread in the precession period which ranges between 1 and 2 ns.

In Fig 5, we plot the $m_x(t)$ trajectories when $V_{VCMA}$ and stress are both turned off abruptly at time $t = t_s$. We continue to monitor the time evolution of $m_x$ until steady state is reached (or approached closely). The four sub-figures in Fig. 5 correspond to the following four situations: (uniaxial) compressive stress along the major axis, tensile stress along the major axis, compressive stress along the minor axis, and tensile stress along the minor axis. Note that steady state is reached when all trajectories end up at either $m_x = -1$ or +1. The former situation represents successful switching and the latter represents failed switching where the failure is caused by thermal noise. The switching error probability is the fraction of trajectories that end up in the wrong state $m_x = +1$. This error probability depends on many factors: $t_s$, $V_{VCMA}$ and the magnitude of the stress. We have kept the value of the stress constant at 50 MPa, but tuned the other two parameters (differently for the four different cases) in an effort to minimize the switching error probability. Our optimization is not sufficiently fine grained in $t_s$ and $V_{VCMA}$ to bring the error probability down to almost zero, which is why the minimum error probability that we were able to attain was ~12%. More fine grained tuning of $t_s$ and $V_{VCMA}$ may allow us to reduce this further. That exercise, however, is beyond the scope of this letter. Undoubtedly a smaller error probability would be preferable, but the 10-20% error probability is still reasonable given the availability of error detection and correction schemes.

An important consideration for a memory cell is that it should have no more than two terminals to maintain high cell density and be compatible with simple cross-bar architecture. In other words, the *same* two terminals should be used to apply a voltage for "read" and "write" operations. The design shown in Fig. 2 accomplishes this. There are only two terminals that are used for both read and write operations. The voltage dropped over the MTJ stack to induce VCMA is labeled $V_{VCMA}$ and that dropped over the piezoelectric to induce stress in the soft layer is labeled $V_{st}$. They are related to the write voltage $V_{WRITE}$ as $V_{VCMA} = R_{MTJ}/(R_{MTJ} + R_{piezo})V_{WRITE}$; $V_{st} = R_{piezo}/(R_{MTJ} + R_{piezo})V_{WRITE}$, where $R_{MTJ}$ is the resistance of the MTJ stack and $R_{piezo}$ is the resistance of piezoelectric film between the two shorted electrodes and the conducting substrate. Since $R_{MTJ}$ is different in the on- and off-states, we have to choose $V_{WRITE}$ such that the lower value of $V_{VCMA}^{low} = R_{MTJ}^{ON}/(R_{MTJ}^{ON} + R_{piezo})V_{WRITE}$ exceeds the voltage required to switch with precessional motion as opposed to thermal activation. Similarly, the lower value of $V_{st}^{low} = R_{piezo}/(R_{MTJ}^{OFF} + R_{piezo})V_{WRITE}$ must exceed the value required to generate the needed stress. Therefore, $V_{WRITE}$ should exceed the greater of $V_{VCMA}^{\min}(1 + R_{piezo}/R_{ON}^{MTJ}), V_{st}^{\min}(1 + R_{OFF}^{MTJ}/R_{piezo})$. Note that this design requires $V_{VCMA}$ and $V_{st}$ to be turned on and off at the same time; they cannot be controlled independently.

In conclusion, we have proposed and analyzed a precessionally switched p-MTJ based memory cell where data is written with VCMA *without any on-chip magnetic field*. The role of the in-plane magnetic field is played by in-plane stress. This approach introduces some additional energy dissipation needed to generate the stress, but that energy overhead is almost negligible [25]. It is a small price to pay for eliminating the on-chip magnetic field.

**Figure captions**

Fig. 1: (a) VCMA switching without an in-plane magnetic field. The magnetization rotates through $90^0$; (b) VCMA switching when an in plane magnetic field is present and the VCMA voltage pulse width is adjusted to one-half of the period of precession around the magnetic field. The magnetization rotates through $180^0$.

Fig. 2: Precessional switching of an MTJ with VCMA and no magnetic field. The write voltage $V_{WRITE}$ is dropped across the piezoelectric ($V_{st}$) and the MTJ ($V_{VCMA}$). The former generates biaxial strain in the elliptical soft layer (compressive along the major axis and tensile along the minor axis, or vice versa, depending on the direction of the poling of the piezoelectric layer and the voltage polarity). The strain acts as an effective in-plane magnetic field around which the out-of-plane magnetization vector begins to precess. By adjusting the $V_{WRITE}$ pulse width to around one-half of the precessional period, the magnetization of the soft layer can be flipped to toggle the resistance of the MTJ between the high and low values. During the read cycle, the amplitude of $V_{WRITE}$ is reduced (and, if necessary, the polarity is reversed) to suppress the VCMA effect. This is a 2-terminal device.

Fig. 3: Depiction of the stress-generated effective magnetic field for various stressing scenarios.

Fig. 4: Plots of (a) the polar angle $\theta$, (b) the azimuthal angle $\phi$, and (c) the out-of-plane component of the magnetization vector $m_x$ as a function of time when uniaxial compressive stress of 50 MPa along the major axis of the elliptical soft layer and a VCMA voltage $V_{VCMA}$ of **-0.6295 V** are turned on abruptly at time $t = 0$. Room temperature (300 K) thermal noise is present. The initial values are $\theta = 90^0$, $\phi = 0^0$ and $m_x = 1$. The different plots are for 1,000 different switching trajectories which are all somewhat different from each other because of thermal noise. These plots are generated from stochastic Landau-Lifshitz-Gilbert simulations.

Fig. 5: Plot of $m_x$ versus time when the VCMA voltage $V_{VCMA}$ and stress are turned off abruptly at time $t = t_s$. Room-temperature thermal noise is present. Note that some trajectories end up at the right state $m_x = -1$ while some end up at the wrong state $m_x = +1$. The percentage of trajectories that end up at the wrong state is the switching error probability. (a) uniaxial compressive stress of 50 MPa along the major axis of the elliptical soft layer; $t_s =$ **0.7 ns** and $V_{VCMA} =$ **-0.6295 V**. The switching error probability at room temperature is **11.9%**; (b) uniaxial tensile stress of 50 MPa along the major axis of the elliptical soft layer; $t_s =$ **0.7 ns** and $V_{VCMA} =$ **-0.45 V**. The switching error probability at room temperature is **18.8%**; (c) uniaxial compressive stress of 50 MPa along the minor axis of the elliptical soft layer; $t_s =$ **0.68 ns** and $V_{VCMA} =$ **-0.6295 V**. The switching error probability at room temperature is **13.7%**; (d) uniaxial tensile stress of 50 MPa along the minor axis of the elliptical soft layer; $t_s =$ **0.75ns** and $V_{VCMA} =$ **-0.45 V**. The switching error probability at room temperature is **11.8%**.

Table I: Parameters used in the simulations (soft layer is Terfenol-D)

| | |
|---|---|
| Surface anisotropy per unit area ($K_{s0}$) | 980 μJ m$^{-2}$ [range 320-980 μJ m$^{-2}$] [12,29,30] |
| VCMA constant ($C$) | 1000 fJ V$^{-1}$ m$^{-1}$ [range 3.3-1150 fJ V$^{-1}$ m$^{-1}$] [3,5,6,15,31] |
| Saturation magnetization ($M_s$) | 8× 10$^5$ A/m [24] |
| Magnetostriction coefficient ($\lambda$) | 600 ppm [24] |
| Gilbert damping ($\alpha$) | 0.1 [32] |
| Major axis of soft layer | 100 nm |
| Minor axis of soft layer | 90 nm |
| Thickness of soft layer | 2 nm |
| Young's modulus of soft layer | 80 GPa |
| Spacer layer thickness | 2 nm |

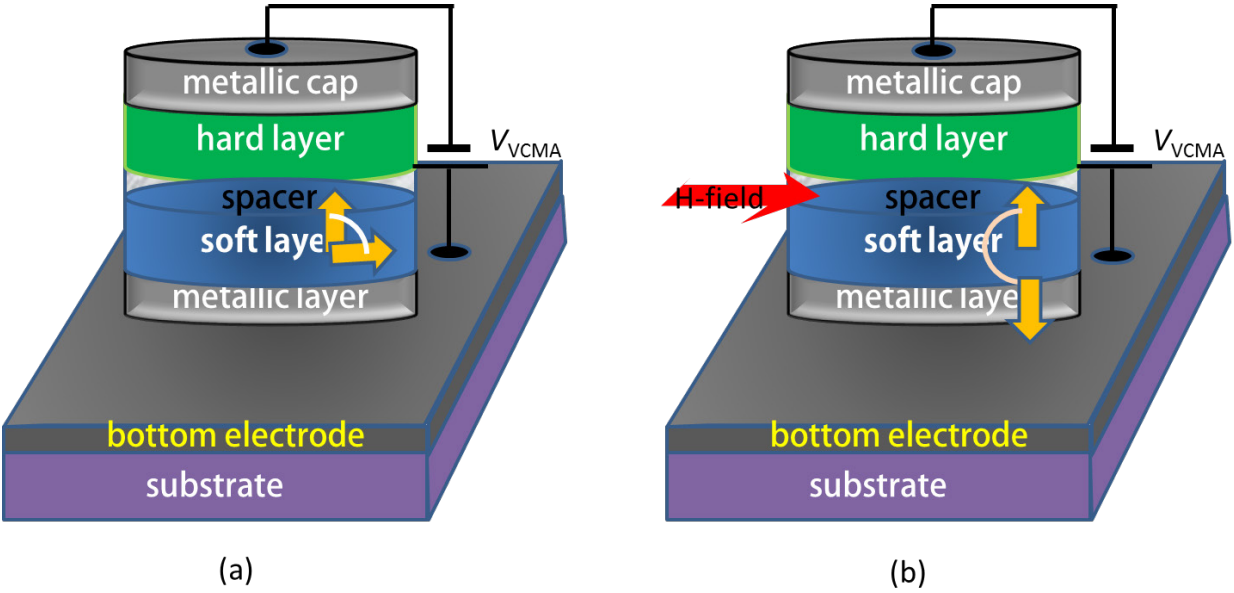

(a) (b)

Fig. 1

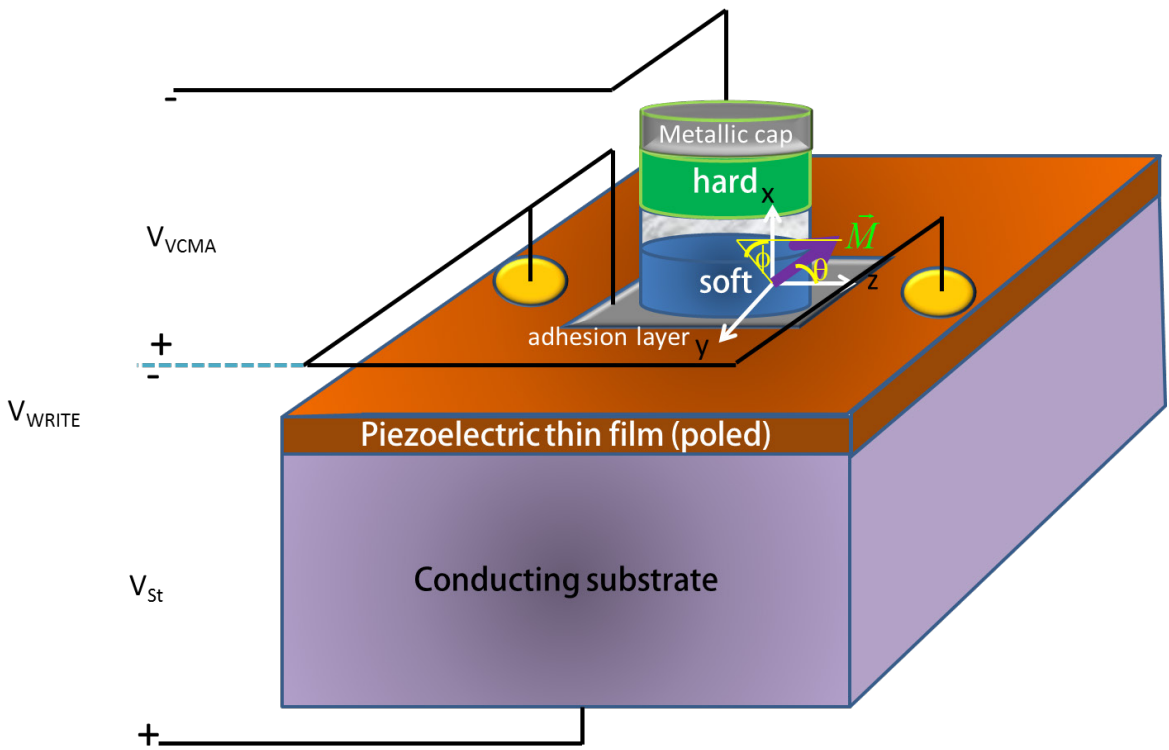

Fig. 2

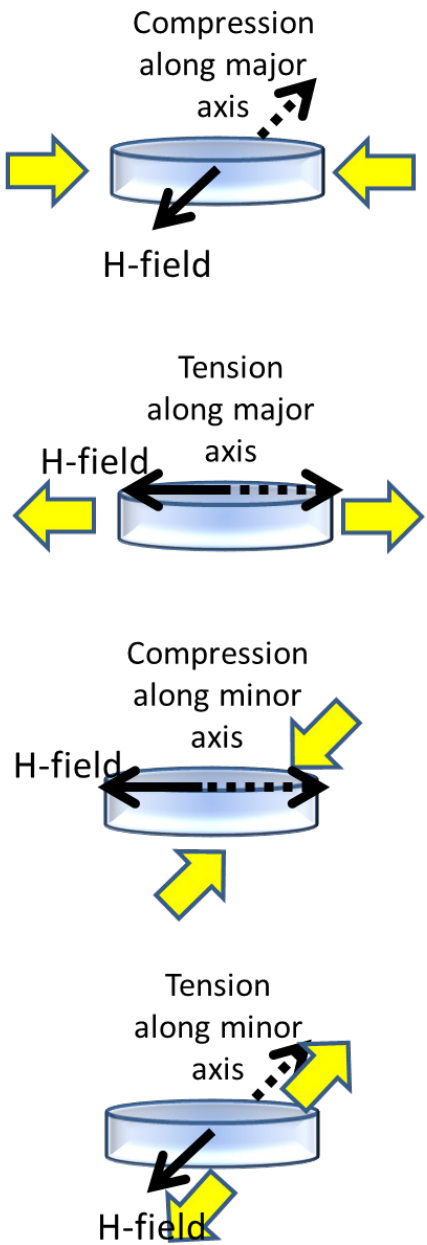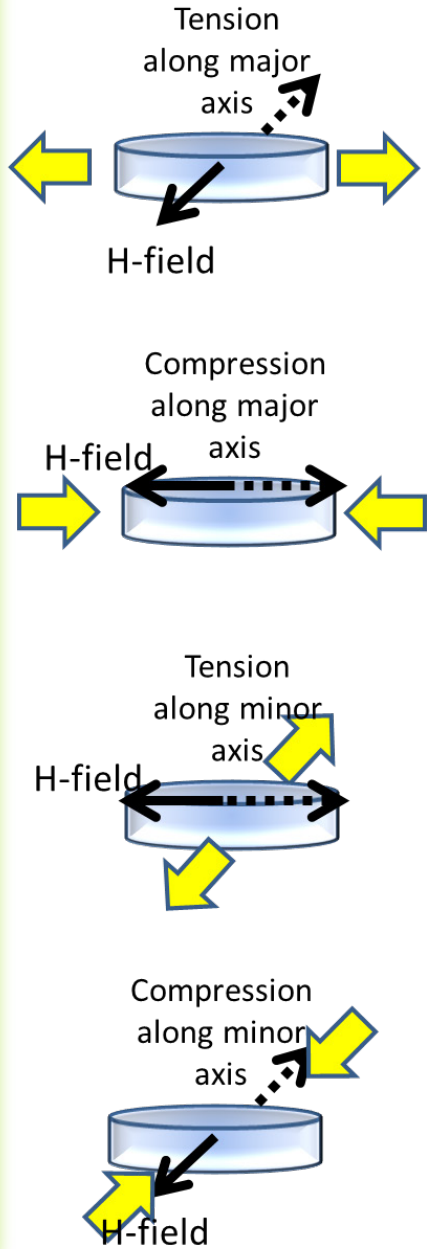

Fig. 3

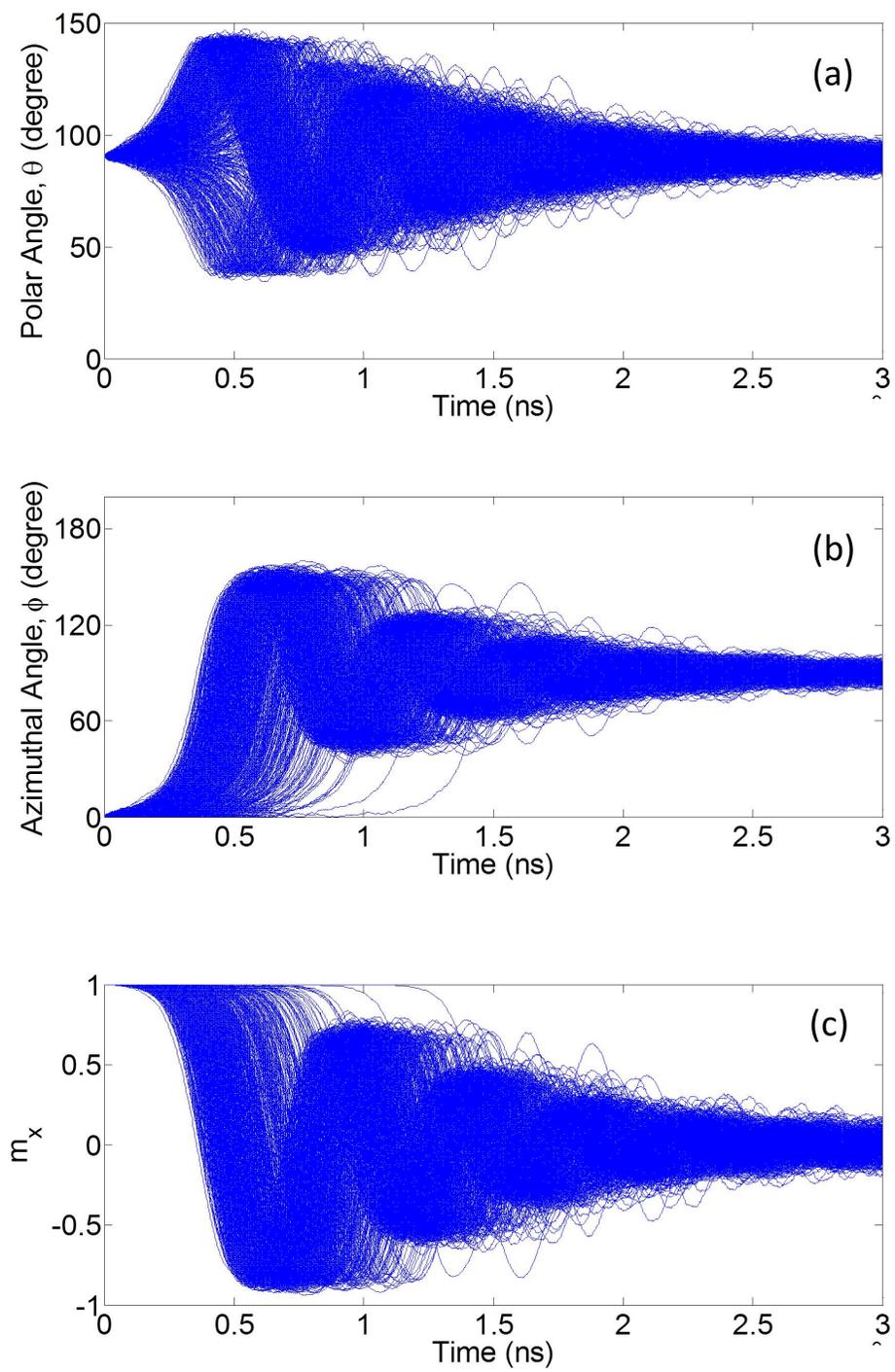

Fig. 4

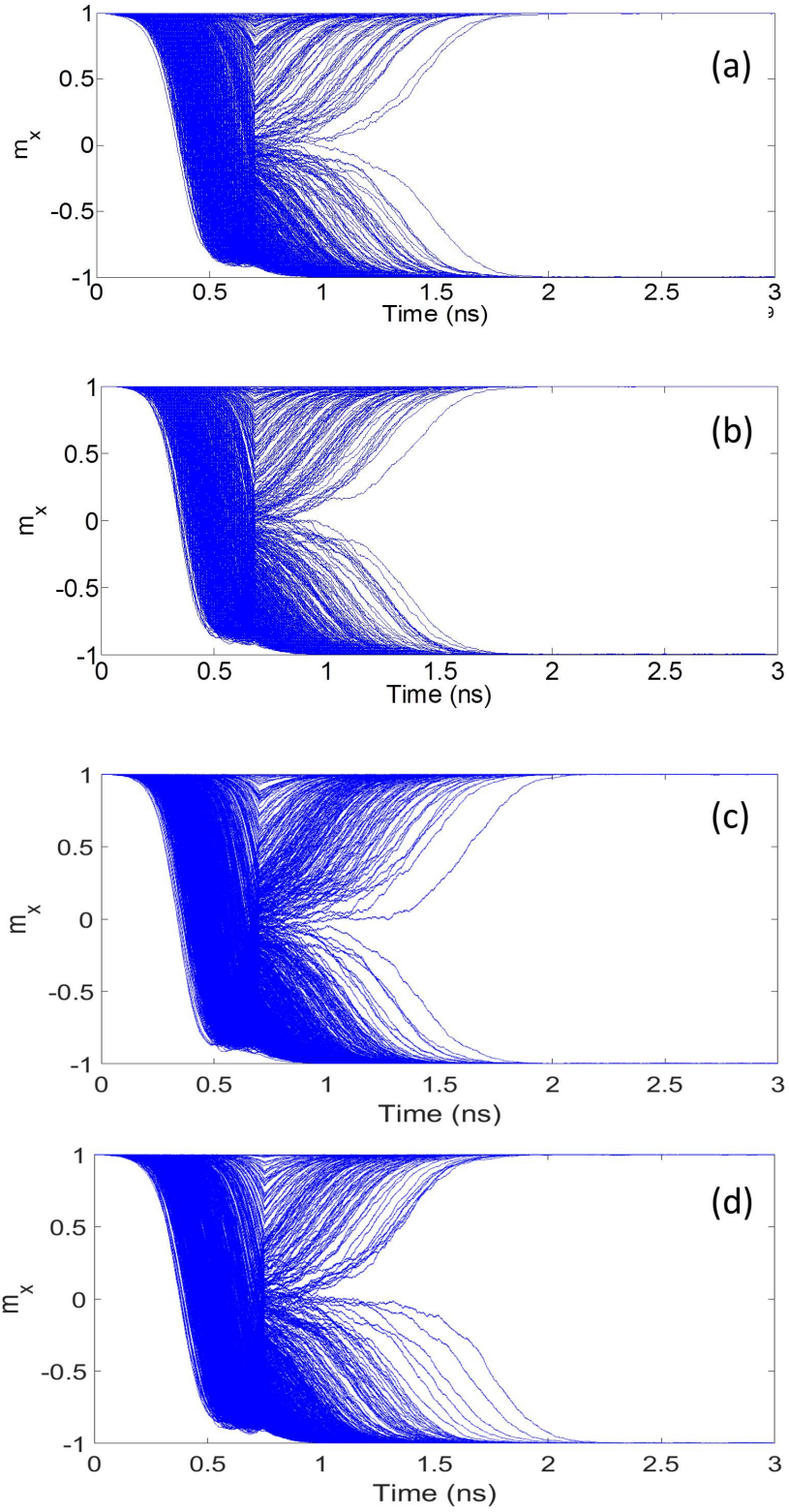

Fig. 5